\documentclass[apl,twocolumn,superscriptaddress,showpacs,nobalancelastpage]{revtex4}

\usepackage{graphicx}
\usepackage{float}

\begin{document}

\title{A few-electron quadruple quantum dot in a closed loop}

\author{Romain Thalineau}
\affiliation{Institut N\'eel, CNRS and Universit\'e Joseph Fourier,38042 Grenoble, France}

\author{Sylvain Hermelin}
\affiliation{Institut N\'eel, CNRS and Universit\'e Joseph Fourier,38042 Grenoble, France}

\author{Andreas D. Wieck}
\affiliation{Lehrstuhl f\"ur Angewandte Festk\"orperphysik, Ruhr-Universit\"at Bochum, Universit\"{a}tsstra\ss e 150, 44780 Bochum, Germany}

\author{Christopher B\"auerle}
\affiliation{Institut N\'eel, CNRS and Universit\'e Joseph Fourier,38042 Grenoble, France}

\author{Laurent Saminadayar}
\affiliation{Institut N\'eel, CNRS and Universit\'e Joseph Fourier,38042 Grenoble, France}

\author{Tristan Meunier}
\affiliation{Institut N\'eel, CNRS and Universit\'e Joseph Fourier,38042 Grenoble, France}

\date{\today}

\begin{abstract}

We report the realization of a quadruple quantum dot device in a square-like configuration where a single electron can be transferred on a closed path free of other electrons. By studying the stability diagrams of this system, we demonstrate that we are able to reach the few-electron regime and to control the electronic population of each quantum dot with gate voltages. This allows us to control the transfer of a single electron on a closed path inside the quadruple dot system. This work opens the route towards electron spin manipulation using spin-orbit interaction by moving an electron on complex paths free of electrons.

\end{abstract}

\maketitle

Controlling the path of a single electron in semiconducting nanostructures is an interesting tool in the context of spin qubits systems. In particular, it opens the route towards the transport of quantum information on a chip and represents a potential strategy to scale up the number of spin qubits in interaction ~\cite{Taylor}. In addition, in presence of spin-orbit interaction, it represents an interesting way to manipulate coherently the spin of a single electron~\cite{Nowack, Salis, Leo}. Recently, fast and efficient single electron transport could be obtained by assisting the transport through a 1D-channel electrostatically defined with surface acoustic waves on AlGaAs heterostructures~\cite{Hermelin, Cambridge}. Nevertheless, such a technique is restricted to displacements on a straight line. To perform more complex displacements, engineering the path of the electron with series of quantum dots is a promising alternative. In this context, it has been demonstrated that topological spin manipulation can be obtained if the electron is transported adiabatically on a closed path under spin-orbit interaction~\cite{Sanjose, Golovach}.

In order to preserve spin information during transport, all other electrons of the heterostructure potentially present along the electron path have to be removed. Therefore all dots have to be emptied and one needs to control them in the so called "few-electron regime". Up to now, only three dots in series have been demonstrated to be tunable ~\cite{Gaudreau, Laird} in the few-electron regime. A triple quantum dot geometry in a star-like configuration has been demonstrated, but the geometry did not allow tunneling between all close-by dots and the few-electron regime was not reached~\cite{Rogge}.

Here we present the first step towards the transfer of a single electron spin on a closed path  using a series of quantum dots in a square-like geometry. We demonstrate that we are able to set the system in the few-electron regime using charge detection techniques and therefore to remove the other electrons along the path. Moreover, the tunability of the four-dot system allows a single electron to be transported along a closed path isolated from the other electrons of the structure.

\begin{figure}
\includegraphics{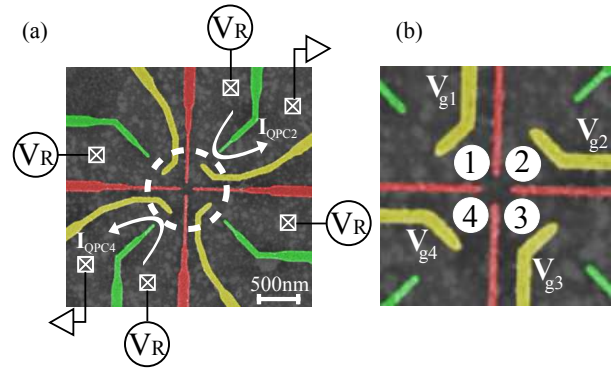}
\caption{(a) SEM image of the device : Ti-Au gates are deposited on the surface of the heterostructure allowing the definition of a quadruple quantum dot. (b) Zoom of the dotted circle of figure \ref{Fig1} (a). Expected position of the four quantum dots with their label.}

\label{Fig1}
\end{figure}

The quantum dot system is fabricated using a GaAs/AlGaAs heterostructure grown by molecular beam epitaxy, with a two-dimensional electron gas (2DEG) 100nm below the surface (density $~2 \times 10^{11}cm^{-2}$  and mobility $~1 \times 10^6V^{-1}s^{-1}$). By applying negative voltages to the metal Schottky gates (Ti-Au) deposited on the surface of the heterostructure, an electrostatic potential confining electrons can be engineered. Figure \ref{Fig1}(a) shows a Scanning Electron Microscopy (SEM) image of the sample used. The pattern of the gates allows us to define a quadruple well potential. Figure \ref{Fig1}(b) shows a zoom image of the sample, in particular the expected position of the four quantum dots and their labels. Each quantum dot of the quadruple well potential is most strongly capacitively coupled to the closest yellow gate and is labeled accordingly. The red gates are used to create a tunnel barrier between the dots while the yellow ones are used to define tunnel barriers with the leads and to control the electrochemical potential of each dot. By modifying the voltage applied to these gates, the electrochemical potential of each dot can be changed, and consequently the electronic population can be controlled.

Quantum point contacts (QPC) are formed with the help of the green gates and used as charge detectors to probe changes in the electronic configuration of the quantum dot system \cite{Field}. Due to the screening of the gates, the coupling of each QPC to its diagonally opposite dot was too small to observe any change in its electronic population. Nevertheless, combining the signal from two QPCs was sufficient to see any charge change in the whole four-dot system. Consequently, only two of them have been used during this experiment, the ones in the upper right ($I_{QPC2}$) and in the lower left ($I_{QPC4}$). They were DC-biased with a voltage $V_R=500\mu$eV, defining also the bias of the four reservoirs. All measurements have been done in a dilution refrigerator with a base temperature around 20mK corresponding to an electronic temperature calibrated to 40mK from weak localization measurement realized in earlier experiments~\cite{YasuPRLPRB1}.

\begin{figure}
	\includegraphics{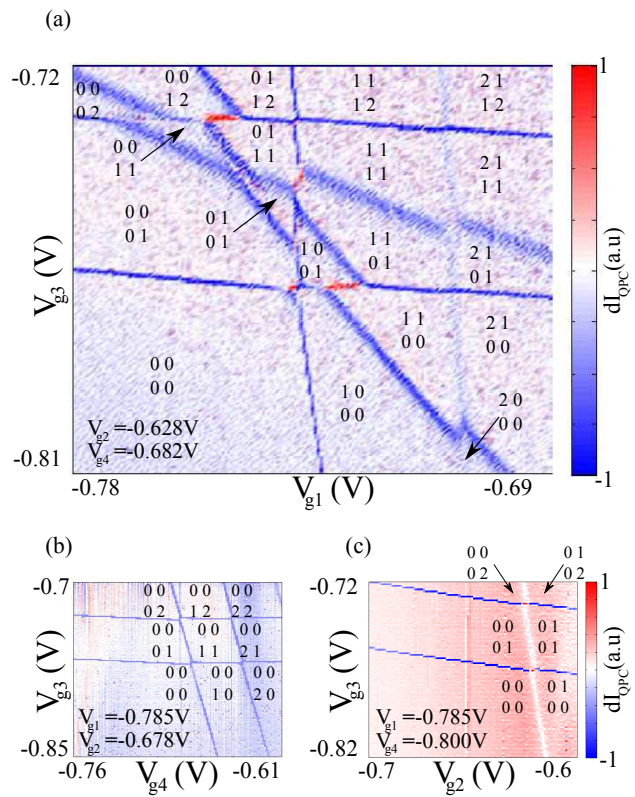}
\caption {(a) Stability diagram of a quadruple quantum dot.  $\frac{dI_{QPC}}{dV}=(\frac{dI_{qpc2}}{dV_{g1}}+\frac{dI_{qpc2}}{dV_{g3}})+(\frac{dI_{qpc4}}{dV_{g1}}+\frac{dI_{qpc4}}{dV_{g3}})$ is plotted as a function of $V_{g1}$ and $V_{g3}$, while $V_{g2}$ and $V_{g4}$ are fixed. The charge label used mimic the geometry of the sample. Upper left label corresponds to the charge contained in the dot strongly coupled to $V_{g1}$, upper right to $V_{g2}$, lower left to $V_{g4}$ and lower right to $V_{g3}$.(b) and (c) Stability diagrams of double dots where the two others are set to unreachable energies. It proofs that the few-electron regime can be reached for each quantum dot. (b) Stability diagram with respect to gate $V_{g3}$ and $V_{g4}$, while $V_{g1}=-0.785$ and $V_{g2}=-0.678$ allowing to say that the dots 1 and 2 are emptied. (c) Stability diagram with respect to gate $V_{g2}$ and $V_{g3}$, while $V_{g1}=-0.785$ and $V_{g4}=-0.8$ allowing to confirm that the dots 1 and 4 are emptied.}
\label{Fig2}
\end{figure}

Figure \ref{Fig2} (a) shows the charge stability diagram with respect to gate voltages $V_{g1}$  and $V_{g3}$  (for $V_{g2}$ and $V_{g4}$ fixed) obtained from analyzing the QPCs current. We have chosen to plot $\frac{dI_{QPC}}{dV}=(\frac{dI_{qpc2}}{dV_{g1}}+\frac{dI_{qpc2}}{dV_{g3}})+(\frac{dI_{qpc4}}{dV_{g1}}+\frac{dI_{qpc4}}{dV_{g3}})$ to emphasize any change of all electronic populations. As expected from the sample geometry, we observe four different types of charge degeneracy lines that we can identify with their slopes with respect to $V_{g1}$  and $V_{g3}$. They delimit Coulomb blockaded regions where the number of electrons in each dot is fixed. Each of these lines correspond to the exchange of one electron between one of the dots and its closest reservoir and their slopes depend on their relative capacitive coupling to the two gates $V_{g1}$ and $V_{g3}$. First, the dot 1 and 3 are  hardly coupled to $V_{g3}$  and $V_{g1}$ respectively. As a consequence, the two corresponding degeneracy lines are the one with the largest (dot 1) and the smallest (dot 3) slope in figure \ref{Fig2}(a). Second, the two remaining degeneracy lines correspond to dot 2 and 4. As expected from the sample geometry, these two dots are coupled almost symmetrically to gates $V_{g1}$ and $V_{g3}$. The slope differences of the corresponding degeneracy lines observed in figure \ref{Fig2}(a) are explained by the geometry of the yellow gates. These gates break the symmetry of the sample. For instance, $V_{g1}$ is more capacitively coupled to dot 2 than to dot 4.  Consequently, the degeneracy line of quantum dot 2 has to exhibit a larger slope than the one of dot 4. One can therefore assign to each charge degeneracy line a corresponding quantum dot by comparing the data of figure \ref{Fig2}(a) with the sample geometry and four quantum dots on a closed loop have been engineered in this way.

Thanks to our identification of the four-dot positions and their corresponding charge degeneracy lines, we can infer the charge configuration of each Coulomb blockade region. The charge label used is defined from the sample geometry (see figure \ref{Fig1}(b)). To understand the indexing, it is convenient to start from the emptied region (lower left part of figure \ref{Fig2}(a)). By increasing voltages $V_{g1}$ or $V_{g3}$, we can add electrons to the four-dot system by crossing one of these charge degeneracy lines. Depending on which charge degeneracy line is crossed, the added electron is labeled accordingly.

Until now, it has been assumed that the lower left part of figure \ref{Fig2}(a) corresponds to the emptied Coulomb blockade region. The charge detection can demonstrate that the proposed labels correspond to an absolute number of charge present in the four-dot system. In figure \ref{Fig2}(a), no degeneracy line is observed for $V_{g1}<-0.74V$ and $V_{g3}<-0.77V$. Due to  the strong capacitive coupling between these two gates and the two dots 1 and 3, one would have expected to observe degeneracy lines in this gate voltage region if the two dots were not emptied. This demonstrates that the dots 1 and 3 have been emptied. On the contrary, we cannot conclude from figure \ref{Fig2}(a) on the population of the dots 2 and 4 due to the small capacitive coupling with gates $V_{g1}$ and $V_{g3}$. To check whether the few-electron regime was reached for dot 2 and dot 4, stability diagrams varying gate voltages $V_{g2}$  and $V_{g4}$  were recorded. Figures \ref{Fig2}(b) and (c) show the QPC response with respect to the gates $V_{g3}$  and $V_{g4}$ ($V_{g2}$). For both cases, the dot chemical potentials have been set such that the dot 1 is empty ($V_{g1} < - 0.785 V$). Such a large negative voltage is also applied to $V_{g2}$($V_{g4}$) in figure \ref{Fig2} (b) (figure \ref{Fig2} (c)) in order to empty the dot 2 (4). In both figures, we demonstrate that no degeneracy line corresponding to dots 2 and 4 are observed, confirming that no electron are in the four-dot system in the bottom left Coulomb blockade region of figure \ref{Fig2}(a).
 
 In addition, one can extract information from figure \ref{Fig2}(a) about the distance between the dots. In this square-like geometry, we expect that the distance between quantum dots to be larger if they are sitting on opposite corners rather than on adjacent corners. The closer the dots are, the larger the inter-dot capacitive coupling is. In the charge stability diagram, this coupling opens a Coulomb gap at the crossings between charge degeneracy lines. Figure \ref{Fig2}(a) clearly shows that the gaps opened at the dot 2-dot 4 and dot 1-dot 3 degeneracy line crossings are much smaller than the other ones confirming the square-like distribution of the four quantum dots expected from the sample geometry. From this set of data, we can therefore demonstrate that by changing the gate voltages $V_{gi}$ (for $i \in \{ 1,2,3,4 \} $, see figure \ref{Fig1} (b)), we are able to remove all the electrons from the quantum dot system, control the injection of a single electron within the four-dot structure as well as its transfer from one dot to the other.

\begin{figure}
\includegraphics{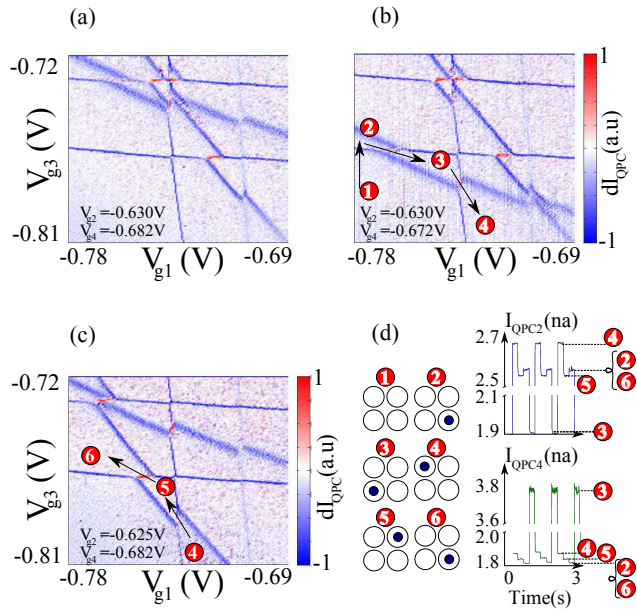}
\caption {(a), (b) and (c) Stability diagram of the quadruple quantum dot for different values of $V_{g2}$ and $V_{g4}$. (b) and (c) Trajectory of the electron in the gate voltage space. d) Position of the electron during the path and the corresponding QPC signal. Each step observed in the QPC currents corresponds to a charge state.}
\label{Fig3}
\end{figure}

The control of the four quantum dot system opens the route towards single electron transport on a closed path. In this last part, we want to give a strategy to perform such an electron displacement. The idea is to keep in close vicinity the chemical potential of dots 1 and 3 and to bring sequentially in this vicinity the chemical potential for dots 2 and 4. By modifying $V_{g2}$ and $V_{g4}$ at the same time, such chemical potential movements can be engineered as demonstrated in the three stability diagrams presented in figures \ref{Fig3}(a), (b) and (c). By changing the voltage applied to one of these two gates, only one charge degeneracy line moves as expected from the sample geometry and demonstrated in figure \ref{Fig3}. Figures \ref{Fig3}(b) and (c) are the two gate voltage configurations used in the transport sequence. On figure \ref{Fig3}(b), the Coulomb blockaded regions {(0,0,0,0)/(1,0,0,0)/(0,0,1,0)/(0,0,0,1)}, where (i,j,k,l) represent the number of electrons in the dots (1,2,3,4), have been set in close vicinity. Starting from the empty region (0,0,0,0) (label 1 in figure \ref{Fig3}(b)), an electron can be loaded into the dot 3 by lowering $|V_{g3}|$ (label 2 in figure \ref{Fig3}(b)). By changing voltages on the two gates $V_{g1}$  and $V_{g3}$ , this electron can then be transferred from dot 3 to dot 1 via two tunnel processes (label 3 and 4 in figure \ref{Fig3}(b)). To complete the closed path, we need to transfer this electron into the dot 2. Consequently the region (0,1,0,0) has to be set in vicinity with (1,0,0,0) and (0,0,1,0). This can be achieved by lowering $|V_{g2}|$ and increasing $|V_{g4}|$ as demonstrated in figure \ref{Fig3}(c). In a similar way, we can then transfer the electron from dot 1 (label 4 in figure \ref{Fig3}(c)) to dot 3 (label 6 in figure \ref{Fig3}(c)) through dot 2 (label 5 in figure \ref{Fig3}(c)). A strategy to repeatedly  displace a single electron on a closed loop is therefore possible. The position of the electron (with respect to the corresponding labels) and the response of the QPC during the repeated transfer are shown in figure \ref{Fig3}(d). Each QPC response step corresponds to a wait time in a stable charge configuration and fast transitions between steps appear whenever the electron transfer takes place. Movement of the gates are faster than the bandwidth of the detector (set to 800Hz). Possible exchange of the electron between dots or with the leads are too fast to be observable with our current set-up.

To be able to use such a transfer to manipulate the spin of the electron, one needs to realize the transfer on time scales faster than the spin decoherence time which is of the order of 10-100ns~\cite{Petta, Bluhm}. Tunneling between the close-by quantum dots needs then to be strong enough. Indeed two competing mechanisms can explain a change of charge configuration when the electron is transferred between two tunnel-coupled dots. First the electron can tunnel directly between the two dots and second it can be replaced in the dot system by an electron coming from the leads. The second scenario requires two energetically allowed tunnel processes, each of them is an electron transfer between one dot and its closest reservoir: the electron tunnels out of the dot initially occupied and an electron from the reservoir tunnels into the dot initially empty. The timescale of the second scenario is set by the coupling of the dots with their connected leads. The results of the slow QPC measurement presented in \ref{Fig3}(d) do not permit to discriminate between these two mechanisms. Moreover, due to the restricted number of gates used to define the dots, we were not able to tune all the tunnel couplings of this system in the regime of a few tens of $\mu$eV. This coupling corresponds to a tunnel timescale inferior to 1ns, which would be at least three orders of magnitude faster than the one between the dots and the leads. Consequently, with such tunnel coupling, we would be able to transfer the electron along this closed path for several turns with a small probability of an exchange with one electron from the reservoirs. A more complete study of the tunneling in this dot configuration is needed to reach the nanosecond transfer of a single electron in a closed loop.


In conclusion, we demonstrated that a tunable four quantum dot system where each dot is located at the corner of a square in the few-electron regime can be engineered and probed with charge detection techniques. It allowed us to transfer a single electron in a closed loop free of electrons on a slow timescale. Addition of gates in the present geometry and coupling to fast electronics should permit to transfer a single electron faster than the spin decoherence and to study topological spin manipulation using spin-orbit interaction.

\begin{acknowledgments}
We acknowledge technical support from the "Poles Electroniques" of the "D\'epartement Nano and MCBT" from the Institut N\'eel as well as Pierre Perrier for technical support. A.D.W. acknowledges expert help from PD Dr. Dirk Reuter and support of the DFG SPP1285 and the BMBF QuaHLRep 01BQ1035. T.M. acknowledges financial support from Marie-Curie ERG 224786. We are grateful to the Nanoscience Foundation of Grenoble, for partial financial support of this work. Devices were fabricated at "Plateforme Technologique Amont" de Grenoble, with the financial support of the "Nanosciences aux limites de la Nano\'electronique" Foundation and CNRS Renatech network.
\end{acknowledgments}

\end{document}